\documentclass[english,aps,pra,manuscript]{revtex4}
\usepackage[T1]{fontenc}
\usepackage[latin9]{inputenc}
\usepackage{amsmath}
\usepackage{amssymb}
\usepackage{graphicx}
\usepackage{esint}

\makeatletter
\@ifundefined{textcolor}{}
{%
 \definecolor{BLACK}{gray}{0}
 \definecolor{WHITE}{gray}{1}
 \definecolor{RED}{rgb}{1,0,0}
 \definecolor{GREEN}{rgb}{0,1,0}
 \definecolor{BLUE}{rgb}{0,0,1}
 \definecolor{CYAN}{cmyk}{1,0,0,0}
 \definecolor{MAGENTA}{cmyk}{0,1,0,0}
 \definecolor{YELLOW}{cmyk}{0,0,1,0}
 }

\@ifundefined{definecolor}{\usepackage{color}}{}

\usepackage{babel}

\makeatother

\usepackage{babel}
\begin{document}

\title{How much time does a measurement take?}

\author{Carlos Alexandre Brasil}

\email{carlosbrasil.physics@gmail.com}

\affiliation{Instituto de Física \textquotedbl{}Gleb Wataghin\textquotedbl{},
Universidade Estadual de Campinas, P.O. Box 6165, 13083-970 Campinas,
São Paulo, Brazil}

\author{L. A. de Castro}

\email{leonardo.castro@usp.br}

\author{R. d. J. Napolitano}

\affiliation{Instituto de Física de São Carlos, Universidade de São Paulo, P.O.
Box 369, 13560-970, São Carlos, SP, Brazil +55 (16) 3373-9820}
\begin{abstract}
We consider the problem of measurement using the Lindblad equation,
which allows the introduction of time in the interaction between the
measured system and the measurement apparatus. We use analytic results,
valid for weak system-environment coupling, obtained for a two-level
system in contact with a measurer (Markovian interaction) and a thermal
bath (non-Markovian interaction), where the measured observable may
or may not commute with the system-environment interaction. Analysing
the behavior of the coherence, which tends to a value asymptotically
close to zero, we obtain an expression for the time of measurement
which depends only on the system-measurer coupling, and which does
not depend on whether the observable commutes with the system-bath
interaction. The behavior of the coherences in the case of strong
system-environment coupling, found numerically, indicates that an
increase in this coupling decreases the measurement time, thus allowing
our expression to be considered the upper limit for the duration of
the process. 
\end{abstract}



\maketitle

\section{Introduction}

In quantum mechanics, the state (or wave function) of a system can
evolve in two distinct ways: unitarily, according to Schrödinger's
equation, when no measurement is being made; or non-unitarily, when
a measurement is made on the system, with the reduction of the wave
function in one of the eigenstates of the observable. \cite{key-1,key-2}
It is exactly in the second case that resides the polemical trait
of the quantum theory, as it makes only statistical predictions about
the results of the measurement. John von Neumann discussed this problem
broadly in his classic book \cite{key-1}, where he admits that the
statistical character of the measurement cannot be omitted. To him,
the measurement involves necessarily the interaction between the system
whose state we wish to determine, and a measuring apparatus whose
state is completely known, so that there will be a transference of
information between the system and the apparatus. Fundamentally, according
to von Neumann, the measurement provides information about the system
indirectly through the apparatus, which, after the interaction, is
in a superposition of states related to the different eigenstates
(and eigenresults) of the main system, then requiring the reduction
postulate to determine the probabilities of obtaining a certain value.
Proceeding with this reasoning, Asher Peres presents a view \cite{key-7}
where the procedure to obtain information from the system, called
\emph{intervention}, is divided in two parts: the \emph{measurement},
when the apparatus interacts with the system and acquires information,
and the \emph{reading} (or \emph{output}), when the result of the
intervention is made known and the reduction occurs, and when we then
obtain the probabilistic information from the diagonal elements of
the density matrix - the populations. Thus, the intrinsically statistical
character of quantum mechanics is related to the reading.

Supposing that the system-measurer interaction can be analysed with
Schrödinger's equation (or, more precisely, with the Liouville-von
Neumann equation)\cite{key-1,key-2}, using a Hamiltonian that takes
into account the main system, the measuring device, and the \emph{Markovian}
interaction between the two, it is found an equation where, to the
Liouvillian - referring to unitary evolutions - it is added a new
term, the Lindbladian - referring to non-unitary processes. This is
Lindblad's equation \cite{key-9,key-8}, originally obtained in a
more general context of quantum dynamical semi-groups \cite{key-10}
for the analysis of irreversible phenomena, but which is used to treat
the measurement process too \cite{key-7,key-19}. In particular, in
Ref. \cite{key-7} Peres cites other works about the derivation of
the Lindblad equation in the measurement context.

When applied to the measurement process, the Lindblad equation allows
something von Neumann did not treat explicitly: to introduce time
in the system-measurer interaction. The result can be known by applying
the reading to the final density operator, thus maintaining the statistical
character of the measurement.

It is our aim here to present results regarding the time evolution
of the measurement process, in a very simple illustration, when the
system being investigated by measurements is not isolated from environmental
perturbations. This work is a natural development of our former papers
\cite{key-13,key-14}, where we considered that the system of interest
is interacting with an environment and a distinct measurement apparatus.
We use the effective Lindbladian \cite{key-9,key-10} description
of the measuring apparatus, whose interaction with the system we assume
as Markovian. However, to treat the noise introduced by the fact that,
during the finite-duration measurement, the system is perturbed by
the environment, we use a non-Markovian Redfield approach. In \cite{key-13},
we developed a method, based on super-operator algebra and Nakajima-Zwanzig
projectors \cite{key-15,key-16} to simplify the treatment of the
\emph{environment} and \emph{principal system + measurement apparatus}
contributions. As we do not include the apparatus as part of the environment,
we end up formulating an unprecedented hybrid description of a noisy
measurement \cite{key-13,key-14}. This method was applied in\cite{key-14}
to analyse the situation where a two-state system (the principal system)
is interacting with an thermal bath (the environment) where, with
some particularizations (that will be explained in Sec. II of the
present communication) we obtained an interesting result, i.e., that
measurements of finite duration performed on an open two-state system
can protect the initial state from a phase-noisy environment, provided
the measured observable does not commute with the perturbing interaction.
The protection is based on the behavior of populations, the diagonal
elements of the density operator. However, there are other elements
whose behavior was not explored in the previous works: the coherences,
the off-diagonal elements of the density operator, closely related
to quantum interferences between the different possible results of
a measurement \cite{key-1,key-2,key-11}. This hiatus is filled with
the present communication.

In this article we present a simple expression for the duration of
the measurement procedure, which is the time the measurement apparatus
must be left interacting with the system until the reading can be
performed within a minimum error margin. The important presupposition
is that the system-measurement interaction can be controlled by the
experimental physicist in the lab so that, with the concepts showed
here, the experimental system can be treated in a form to better interpret
its results. To this end, we will employ the analytical solutions
found in Ref. \cite{key-14}, valid for weak system-environment interaction,
through the modulus of the coherences, which tend to a value asymptotically
close to zero after a certain time. We will consider the time for
the system-measurer interaction to end as the instant when the modulus
of the coherence reaches a certain small fraction of its original
value.

The analytic expression presents some interesting features: it does
not depend on whether the measured observable commutes with the system-environment
interaction, it does not depend on the initial conditions, it does
not depend on the system-environment coupling (even though this is
valid only for weak couplings), and depends only on the system-measurer
coupling. As expected, the stronger the system-measurer coupling (\emph{i.e.},
the more intense the measurement), the less time is necessary to complete
the reading.

Employing a numerical method that allows the analysis of cases with
strong system-environment couplings, we found that the time of measurement
decreases as the coupling increases, for the same system-measurer
interaction. Therefore, our expression, obtained from analytical solutions,
is an upper limit for the time of measurement.

In our treatment, we start from an equation for the evolution of the
total density operator which, before tracing out of the environmental
degrees of freedom, involves a unitary evolution of the interaction
between the system and its environment, together with a non-unitary
Lindbladian evolution of the interaction between the system and the
measuring device. Therefore, the degrees of freedom of the device
might be thought of as already having been traced out of the formulation,
so that the only tracing out left regards the degrees of freedom describing
the non-Markovian noise.

As showed in the beginning of this introduction, the quantum measurement
theory is a wide topic for studies, with several and distinct approaches.
However, the analysis of the duration of the measurement presented
here is unprecedented and entirely new.

In our studies, we will not consider the process of reduction of the
wave function, which displays the statistical character of the intervention.
There are interpretations of quantum mechanics \cite{key-3,key-4,key-5}
where the reduction is deemed inexistent. A recent review of this
subject can be found in Ref. \cite{key-6}.

This article is structured as follows: in Sec. II we deduce the analytical
expression of the measurement time; in Sec. III we analyse the validity
of this expression for different system-environment coupling intensities
using a numerical method; and we conclude in Sec. IV.

\section{Measurement Time}

\subsection{The hybrid master equation}

To derive the hybrid master equation in Ref. \cite{key-13}, we have
considered a main system $S$ which, during the measurement process,
is interacting with an environment $B$. Their evolution is governed
by the Lindblad equation \cite{key-9}, 
\begin{equation}
\frac{d}{dt}\hat{\rho}_{SB}\left(t\right)=-\frac{i}{\hbar}\left[\hat{H},\hat{\rho}_{SB}\left(t\right)\right]+\underset{j}{\sum}\left(\hat{L}_{j}\hat{\rho}_{SB}\left(t\right)\hat{L}_{j}^{\dagger}-\frac{1}{2}\left\{ \hat{L}_{j}^{\dagger}\hat{L}_{j},\hat{\rho}_{SB}\left(t\right)\right\} \right),\label{lindgeral}
\end{equation}
 where $\hat{\rho}_{SB}\left(t\right)$ is the total density operator,
$\hat{H}$ is the total Hamiltonian and the $\hat{L}_{j}$ are the
Lindblad operators that act only on the system. The first term on
the right-hand side acting on $\hat{\rho}_{SB}\left(t\right)$ is
the Liouvillian superoperator, which accounts for the unitary portion
of the propagation, while the second term, the Lindbladian superoperator,
represents the Markovian measurement dynamics.

In the Liouvillian term of Eq. (\ref{lindgeral}), the total Hamiltonian
can be split in terms $\hat{H}_{S}$ and $\hat{H}_{B}$, which act
only on $S$ and $B$, respectively, and an interaction term $\hat{H}_{SB}$:

\[
\hat{H}=\hat{H}_{B}+\hat{H}_{SB}+\hat{H}_{S}.
\]
 To model the non-Markovian noise, we suppose that the interaction
term that can be decomposed in:

\begin{eqnarray}
\hat{H}_{SB} & = & \sum_{k}\hat{S}_{k}\hat{B}_{k},\label{defHSB}
\end{eqnarray}
 where the $\hat{S}_{k}$ operate only on the system $S$, and $\hat{B}_{k}$,
only on the environment $B$. The form of the interaction given by
Eq. (\ref{defHSB}) is capable of describing both \emph{amplitude-damping}
and \emph{phase-damping} quantum channels \cite{key-11}.

The Lindbladian term of Eq. (\ref{lindgeral}) will act solely on
the Hilbert space of the system $S$, since we are interested in measuring
system observables only. Using this information about which parts
of each superoperator act on which Hilbert spaces, the right-hand
side of the Lindblad equation (\ref{lindgeral}) can be split in two
commuting superoperators $\hat{\hat{B}}$ and $\hat{\hat{S}}$ that
act only on the environment or the system, respectively, 
\begin{eqnarray}
\hat{\hat{B}}\hat{X} & = & -\frac{i}{\hbar}\left[\hat{H}_{B},\hat{X}\right],\label{defB}
\end{eqnarray}
 
\begin{eqnarray*}
\hat{\hat{S}}\hat{X} & = & -\frac{i}{\hbar}\left[\hat{H}_{S},\hat{X}\right]+\underset{j}{\sum}\left(\hat{L}_{j}\hat{X}\hat{L}_{j}^{\dagger}-\frac{1}{2}\left\{ \hat{L}_{j}^{\dagger}\hat{L}_{j},\hat{X}\right\} \right),
\end{eqnarray*}
 and an interaction superoperator $\hat{\hat{F}}$, that acts on both
Hilbert spaces: 
\begin{eqnarray}
\hat{\hat{F}}\hat{X} & = & -\frac{i}{\hbar}\left[\hat{H}_{SB},\hat{X}\right].\label{defF}
\end{eqnarray}

From this dynamical equation, we have employed the Nakajima-Zwanzig
projector superoperator $\hat{\hat{P}}$ \cite{key-15,key-16}, defined
as 
\begin{eqnarray}
\hat{\hat{P}}\hat{X}\left(t\right) & = & \hat{\rho}_{B}\left(t_{0}\right)\otimes\mathrm{Tr}_{B}\left\{ \hat{X}\left(t\right)\right\} ,\label{defP}
\end{eqnarray}
 to obtain the hybrid master equation, 
\begin{eqnarray}
\frac{d}{dt}\left[\hat{\hat{P}}\hat{\alpha}\left(t\right)\right] & = & \int_{0}^{t}dt^{\prime}\,\left[\hat{\hat{P}}\hat{\hat{G}}\left(t\right)\hat{\hat{G}}\left(t^{\prime}\right)\hat{\hat{P}}\hat{\alpha}\left(t\right)\right],\label{eqFINAL}
\end{eqnarray}
 where 
\begin{eqnarray}
\hat{\alpha}\left(t\right) & \equiv & e^{-\hat{\hat{S}}t-\hat{\hat{B}}t}\hat{\rho}_{SB}\left(t\right),\label{defalfa}
\end{eqnarray}
 and 
\begin{eqnarray}
\hat{\hat{G}}\left(t\right) & \equiv & e^{-\hat{\hat{S}}t-\hat{\hat{B}}t}\hat{\hat{F}}e^{\hat{\hat{S}}t+\hat{\hat{B}}t}.\label{defG}
\end{eqnarray}

To obtain the Eq. (\ref{eqFINAL}), it is important to emphasize that
$\hat{\hat{P}}\hat{\hat{G}}\left(t\right)\hat{\hat{G}}\left(t^{\prime}\right)\hat{\hat{P}}\hat{\alpha}\left(0\right)=0$
- see Ref. \cite{key-13}.

Finally, the reduced density operator $\hat{\rho}_{S}\left(t\right)$,
which gives the relevant information about the state of the system,
can be found from $\hat{\alpha}\left(t\right)$ as defined in Eq.
(\ref{defalfa}): 
\[
\hat{\rho}_{S}\left(t\right)\equiv\mathrm{Tr}_{B}\left\{ \hat{\rho}_{SB}\left(t\right)\right\} =e^{\hat{\hat{S}}t}\mathrm{Tr}_{B}\left\{ \hat{\alpha}\left(t\right)\right\} .
\]

\subsection{The specific solutions}

In Ref. \cite{key-14} we have solved the master equation (\ref{eqFINAL})
for two different types of measurements. In both cases, we used the
following system and environmental Hamiltonians: 
\[
\hat{H}_{S}=\hbar\omega_{0}\hat{\sigma}_{z},
\]
 
\begin{equation}
\hat{H}_{B}=\hbar\underset{k}{\sum}\omega_{k}\hat{b}_{k}^{\dagger}\hat{b}_{k},\label{Hb}
\end{equation}
 together with a \emph{phase-damping }interaction\emph{ }\cite{key-11},
that is characterized by the following operators in Eq. (\ref{defHSB}):
\[
\begin{cases}
\hat{S}_{k} & =\hbar\hat{\sigma}_{z},\\
\hat{B}_{k} & =g_{k}\hat{b}_{k}^{\dagger}+g_{k}^{*}\hat{b}_{k},
\end{cases}
\]
 where the $\hat{\sigma}_{\alpha}\:,\:\alpha=x,z$ are the Pauli matrices

\begin{equation}
\hat{\sigma}_{z}=\left(\begin{array}{cc}
1 & 0\\
0 & -1
\end{array}\right),\:\hat{\sigma}_{x}=\left(\begin{array}{cc}
0 & 1\\
1 & 0
\end{array}\right),
\end{equation}
 $\omega_{0}$ and the $\omega_{k}$ are real constants, $\hat{b}_{k}$
and $\hat{b}_{k}^{\dagger}$ are the annihilation and creation bosonic
operators, and the $g_{k}$ are complex coefficients. The latter are
constrained by an Ohmic spectral density,

\begin{equation}
J\left(\omega\right)\equiv\sum_{k}\left|g_{k}\right|^{2}\delta\left(\omega_{k}-\omega\right)=\eta\omega e^{-\frac{\omega}{\omega_{c}}},\label{eq:OhmicSD}
\end{equation}
 where $\eta\ge0$ is the constant that gives the strength of the
coupling between the system and its environment, and $\omega_{c}\geqslant0$
is the cutoff frequency. The initial state of the environment is given
by:

\begin{equation}
\hat{\rho}_{B}=\frac{1}{Z_{B}}\underset{k}{\prod}e^{-\hbar\beta\omega_{k}\hat{b}_{k}^{\dagger}\hat{b}_{k}},\, Z_{B}=\underset{l}{\prod}\frac{1}{1-e^{-\hbar\beta\omega_{l}}},\label{rBtermico}
\end{equation}
 where $\beta=\left(kT\right)^{-1}$ represents the initial temperature
of the bath.

Then, considering the cases of measuring observables $\hat{L}=\lambda\hat{\sigma}_{z}$
and $\hat{L}=\lambda\hat{\sigma}_{x}$, we have found the solutions
of Eq. (\ref{eqFINAL}), valid for weak system-environment interaction
$\eta$, shown in the next two sections. There, the $\rho_{ij}\left(t\right)$,
$i,j=1,2$, are the matrix elements of the reduced density operator
$\hat{\rho}_{S}\left(t\right)$, and the upper indices in brackets
in Eq. (\ref{roxfinal}) indicate the basis in which the matrix elements
must be taken: the initial conditions are taken from the eigenbasis
of $\hat{\sigma}_{z}$, $\left\{ \left|+\right\rangle ,\left|-\right\rangle \right\} $,
but the final answers are written in the eigenbasis of the measurement
$\lambda\hat{\sigma}_{x}$, $\left\{ \left|+\right\rangle _{x},\left|-\right\rangle _{x}\right\} $,
where 
\[
\begin{cases}
\left|+\right\rangle _{x} & =\frac{\left|+\right\rangle +\left|-\right\rangle }{\sqrt{2}},\\
\left|-\right\rangle _{x} & =\frac{\left|+\right\rangle -\left|-\right\rangle }{\sqrt{2}}.
\end{cases}
\]

\subsubsection{The case of $\hat{L}=\lambda\hat{\sigma}_{z}$ and $T\neq0$}

We have found the general solutions

\begin{equation}
\begin{cases}
\rho_{11}\left(t\right) & =\rho_{11}\left(0\right),\\
\rho_{12}\left(t\right) & =\rho_{12}\left(0\right)\left[\frac{\Gamma\left(\frac{1}{\omega_{c}\beta\hbar}+i\frac{t}{\beta\hbar}\right)\Gamma\left(\frac{1}{\omega_{c}\beta\hbar}-i\frac{t}{\beta\hbar}\right)}{\Gamma^{2}\left(\frac{1}{\omega_{c}\beta\hbar}\right)}\frac{\Gamma\left(\frac{1}{\omega_{c}\beta\hbar}+1+i\frac{t}{\beta\hbar}\right)\Gamma\left(\frac{1}{\omega_{c}\beta\hbar}+1-i\frac{t}{\beta\hbar}\right)}{\Gamma^{2}\left(\frac{1}{\omega_{c}\beta\hbar}+1\right)}\right]^{2\eta}e^{-2\lambda^{2}t}e^{i2\omega_{0}t}.
\end{cases}\label{solz}
\end{equation}
 In particular, the expression for the coherence was found after solving
the following general integral: 
\begin{equation}
\rho_{12}\left(t\right)=\rho_{12}\left(0\right)\exp\left[-4\eta\int_{0}^{t}dt'\int_{0}^{\infty}d\omega e^{-\frac{\omega}{\omega_{c}}}\sin\left(\omega t'\right)\coth\left(\frac{\beta\hbar\omega}{2}\right)\right]e^{-2\lambda^{2}t}e^{i2\omega_{0}t}\label{coerzint}
\end{equation}

\subsubsection{The case of $\hat{L}=\lambda\hat{\sigma}_{x}$, $T=0$, and $\omega_{0}=0$}

In this case, the particularizations $T=0$ and $\omega_{0}=0$ were
necessary to find the analytical solutions 
\begin{equation}
\begin{cases}
\rho_{11}^{\left(x\right)}\left(t\right)= & \frac{1}{2}+\mathrm{Re}\left\{ \rho_{12}^{\left(z\right)}\left(0\right)\right\} e^{-8\eta\lambda^{2}g_{0}t}e^{4\eta\lambda^{2}\left[A_{-}\left(t\right)-B_{-}\left(t\right)\right]},\\
\rho_{12}^{\left(x\right)}\left(t\right)= & \frac{2\rho_{11}^{\left(z\right)}\left(0\right)-1}{2}e^{-2\lambda^{2}t}-i\mathrm{Im}\left\{ \rho_{12}^{\left(z\right)}\left(0\right)\right\} e^{-2\lambda^{2}t}e^{8\eta\lambda^{2}g_{0}t}e^{-4\eta\lambda^{2}\left[A_{+}\left(t\right)+B_{+}\left(t\right)\right]},
\end{cases}\label{roxfinal}
\end{equation}
 where 
\[
\begin{cases}
A_{+}\left(t\right) & \equiv\int_{0}^{t}e^{2\lambda^{2}t'}g_{1}\left(t'\right)dt',\\
A_{-}\left(t\right) & \equiv\int_{0}^{t}e^{-2\lambda^{2}t'}g_{1}\left(t'\right)dt',\\
B_{+}\left(t\right) & \equiv\int_{0}^{t}e^{2\lambda^{2}t'}g_{2}\left(t'\right)dt',\\
B_{-}\left(t\right) & \equiv\int_{0}^{t}e^{-2\lambda^{2}t'}g_{2}\left(t'\right)dt',
\end{cases}
\]
 and

\[
\begin{cases}
g_{0} & =\int_{0}^{\infty}d\omega\frac{\omega}{4\Omega^{2}+\omega^{2}}e^{-\frac{\omega}{\omega_{c}}},\\
g_{1}\left(t\right) & =2\int_{0}^{\infty}d\omega\frac{\omega}{4\Omega^{2}+\omega^{2}}e^{-\frac{\omega}{\omega_{c}}}\cos\left(\omega t\right),\\
g_{2}\left(t\right) & =\frac{1}{\Omega}\int_{0}^{\infty}d\omega\frac{\omega^{2}}{4\Omega^{2}+\omega^{2}}e^{-\frac{\omega}{\omega_{c}}}\sin\left(\omega t\right),
\end{cases}
\]
where

\begin{eqnarray*}
\Omega & \equiv & \sqrt{\lambda^{4}-4\omega_{0}^{2}}.
\end{eqnarray*}

\subsection{Finding the measurement time}

As we are dealing with the matrix elements of a density operator,
the populations will provide probabilities related to different possible
outcomes. We will consider here, in order to establish a criterion
for the duration of the measurement, the behavior of the coherences.
As it can be seen from the graphs of the moduli of the coherences
in both cases, they tend to a value asymptotically close to zero after
a short period of time (Fig. 1). Hence, the problem consists in finding
a \emph{simple} expression for the time when $\left|\rho_{12}^{\left(\alpha\right)}\left(t\right)\right|,\:\alpha=x,z$,
equals a fraction $f$ of its original value, \emph{i.e.}

\begin{equation}
\left|\rho_{12}^{\left(\alpha\right)}\left(t_{M}\right)\right|=f\left|\rho_{12}^{\left(\alpha\right)}\left(0\right)\right|,\label{problema}
\end{equation}
 where $0<f<1$ for a certain time $t_{M}$. The non-trivial forms
of Eqs. (\ref{coerzint}) and (\ref{roxfinal}) prevent the exact
solution of Eq. (\ref{problema}). Thus, we have to approximate the
expression by means of series expansions, considering $t_{M}\ll1$,
which is justified from the behavior of the two expressions. As both
cases involve exponentials,

\begin{equation}
\left|\rho_{12}\left(t\right)\right|\propto e^{F\left(t\right)},
\end{equation}
 where the $\propto$ signal includes exponentials with linear arguments
and $F\left(t\right)$ is a function whose form depends on the situation
considered, our approach consists in expanding $F\left(t\right)$
in a power series considering terms up to the first order in $t:$

\begin{eqnarray*}
F\left(t\right) & \simeq & F\left(0\right)+F'\left(0\right)t,
\end{eqnarray*}
 replacing the expansion in the argument of the exponential 
\begin{equation}
\left|\rho_{12}\left(t\right)\right|\propto e^{F\left(0\right)+F'\left(0\right)t},\label{coeraproxgeral}
\end{equation}
 and solving Eq. (\ref{problema}) using Eq. (\ref{coeraproxgeral}).

\subsubsection{The case of $\hat{L}=\lambda\hat{\sigma}_{z}$ and $T\neq0$}

Now, we consider Eq. (\ref{coerzint}). The modulus of the coherence
is:

\[
\left|\rho_{12}\left(t\right)\right|=\left|\rho_{12}\left(0\right)\right|\exp\left[-4\eta\int_{0}^{t}dt'\int_{0}^{\infty}d\omega e^{-\frac{\omega}{\omega_{c}}}\sin\left(\omega t'\right)\coth\left(\frac{\beta\hbar\omega}{2}\right)\right]e^{-2\lambda^{2}t},
\]
 so that Eq. (\ref{problema}) becomes

\begin{eqnarray*}
\left|\rho_{12}\left(0\right)\right|\exp\left[-4\eta\int_{0}^{t_{M}}dt'\int_{0}^{\infty}d\omega e^{-\frac{\omega}{\omega_{c}}}\sin\left(\omega t'\right)\coth\left(\frac{\beta\hbar\omega}{2}\right)\right]e^{-2\lambda^{2}t_{M}} & = & f\left|\rho_{12}\left(0\right)\right|,
\end{eqnarray*}
 or, simplifying, 
\begin{eqnarray*}
\exp\left[-4\eta\int_{0}^{t_{M}}dt'\int_{0}^{\infty}d\omega e^{-\frac{\omega}{\omega_{c}}}\sin\left(\omega t'\right)\coth\left(\frac{\beta\hbar\omega}{2}\right)\right]e^{-2\lambda^{2}t_{M}} & = & f.
\end{eqnarray*}
 We apply the methodology of the expansion of the argument of the
first integral defined by the function

\begin{equation}
F\left(t\right)\equiv-4\eta\int_{0}^{t}dt'\int_{0}^{\infty}d\omega e^{-\frac{\omega}{\omega_{c}}}\sin\left(\omega t'\right)\coth\left(\frac{\beta\hbar\omega}{2}\right).
\end{equation}
 So, up to the first order in $t,$ 
\begin{equation}
e^{-2\lambda^{2}t_{M}}=f,
\end{equation}
 and the expression for $t_{M}$ becomes

\begin{equation}
t_{M}=-\frac{1}{2\lambda^{2}}\ln\left(f\right),\label{tempoMz}
\end{equation}
 where $f<1$.

\subsubsection{The case of $\hat{L}=\lambda\hat{\sigma}_{x}$, $T=0$, and $\omega_{0}=0$}

Now we consider the second of Eqs. (\ref{roxfinal}). The square modulus
of the coherence becomes: 
\begin{equation}
\left|\rho_{12}^{\left(x\right)}\left(t\right)\right|^{2}=\left[\frac{2\rho_{11}^{\left(z\right)}\left(0\right)-1}{2}\right]^{2}e^{-4\lambda^{2}t}+\mathrm{Im}\left\{ \rho_{12}^{\left(z\right)}\left(0\right)\right\} ^{2}e^{-4\lambda^{2}t}e^{16\eta\lambda^{2}g_{0}t}e^{-8\eta\lambda^{2}\left[A_{+}\left(t\right)+B_{+}\left(t\right)\right]}.
\end{equation}
 Developing Eq. (\ref{problema}), we find:

\begin{eqnarray*}
\frac{R_{0}}{N_{0}}e^{-4\lambda^{2}t_{M}}+\frac{I_{0}}{N_{0}}e^{-4\lambda^{2}t_{M}}e^{16\eta\lambda^{2}g_{0}t_{M}}e^{-8\eta\lambda^{2}\left[A_{+}\left(t_{M}\right)+B_{+}\left(t_{M}\right)\right]} & = & f^{2},
\end{eqnarray*}
 where

\begin{eqnarray*}
R_{0} & = & \left[\frac{2\rho_{11}^{\left(z\right)}\left(0\right)-1}{2}\right]^{2},\\
I_{0} & = & \mathrm{Im}\left\{ \rho_{12}^{\left(z\right)}\left(0\right)\right\} ^{2},\\
N_{0} & = & R_{0}+I_{0}.
\end{eqnarray*}
 For the sake of simplicity, we define:

\begin{eqnarray*}
k_{1} & = & \frac{R_{0}}{N_{0}},\\
k_{2} & = & \frac{I_{0}}{N_{0}},
\end{eqnarray*}
 for the main equation of our problem to become:

\begin{equation}
k_{1}e^{-4\lambda^{2}t_{M}}+k_{2}e^{-4\lambda^{2}t_{M}}e^{16\eta\lambda^{2}g_{0}t_{M}}e^{-8\eta\lambda^{2}\left[A_{+}\left(t_{M}\right)+B_{+}\left(t_{M}\right)\right]}=f^{2}.\label{principal}
\end{equation}
 We apply the expansion over the last exponential of the second term
on left-hand side,

\begin{equation}
F\left(t\right)=-8\eta\lambda^{2}\left[A_{+}\left(t\right)+B_{+}\left(t\right)\right],
\end{equation}
 then, we have in Eq. (\ref{principal}):

\begin{eqnarray*}
k_{1}e^{-4\lambda^{2}t_{M}}+k_{2}e^{-4\lambda^{2}t_{M}}e^{16\eta\lambda^{2}g_{0}t_{M}}e^{-8\eta\lambda^{2}\left(2g_{0}t_{M}\right)} & = & f^{2}\Rightarrow\\
\Rightarrow k_{1}e^{-4\lambda^{2}t_{M}}+k_{2}e^{-4\lambda^{2}t_{M}} & = & f^{2}\Rightarrow\\
\Rightarrow\frac{k_{1}+k_{2}}{f^{2}} & = & e^{4\lambda^{2}t_{M}}.
\end{eqnarray*}
 Therefore, according to the definitions of $k_{1}$ and $k_{2}$,

\begin{equation}
t_{M}=-\frac{1}{2\lambda^{2}}\ln\left(f\right),\label{tempoMx}
\end{equation}
 keeping in mind that $f<1$. This expression is identical to Eq.
(\ref{tempoMz}), found in the previous item.

Moreover, it is possible to rewrite the second of Eqs. (\ref{roxfinal})
in the form:

\begin{eqnarray*}
\rho_{12}^{\left(x\right)}\left(t\right) & = & \frac{2\rho_{11}^{\left(z\right)}\left(0\right)-1}{2}e^{-2\lambda^{2}t}-i\mathrm{Im}\left\{ \hat{\rho}_{12}^{\left(z\right)}\left(0\right)\right\} \exp\left\{ -2\lambda^{2}t\right\} \\
 &  & \times\exp\left\{ -8\eta\lambda^{2}\int_{0}^{\infty}\mathrm{d}\omega\int_{0}^{t}\mathrm{d}t^{\prime}e^{-\omega/\omega_{c}}\omega\frac{e^{2\lambda^{2}t^{\prime}}\cos\left(\omega t^{\prime}\right)-1}{4\lambda^{2}+\omega^{2}}\right\} \\
 &  & \times\exp\left\{ -4\eta\int_{0}^{\infty}\mathrm{d}\omega\int_{0}^{t}\mathrm{d}t^{\prime}e^{-\omega/\omega_{c}}\omega^{2}\frac{e^{2\lambda^{2}t}}{4\lambda^{2}+\omega^{2}}\sin\left(\omega t^{\prime}\right)\right\} .
\end{eqnarray*}

As we are dealing with small perturbations caused by the environment,
it is safe to assume that the exact measurement time will be much
shorter than the typical decoherence time, so that, in the time periods
we are dealing with in the integrals, $t_{M}\ll\omega_{C}^{-1}$.
Therefore, we can consider that the $\omega t^{\prime}$ in the integrals
is close to zero, thus leading to the approximations $\cos\left(\omega t^{\prime}\right)\approx1$
and $\sin\left(\omega t\right)\approx\omega t$, and guaranteeing
the non-negativity of the two integrands during the characteristic
time of the measurement. We will have, therefore, negative numbers
multiplying the coupling constant, which shows that an increase in
the coupling with the environment increases the speed with which the
measurement is performed.

\subsection{Final expression}

We have found that, regardless of the observable measured, $\hat{L}=\lambda\hat{\sigma}_{x}$
or $\hat{L}=\lambda\hat{\sigma}_{z}$ (even though some particularizations
- $\omega_{0}=0$, $T=0$ - were made in the first case), we have
the same expression for the time of measurement:

\begin{equation}
t_{M}=-\frac{1}{2\lambda^{2}}\ln\left(f\right).\label{tm}
\end{equation}
 This expression does not depend on the system-environment coupling,
$\eta$, nor on initial conditions, but only on the system-measurer
coupling $\lambda$. However, it also depends on the threshold constant
$f$, which is, for the moment, arbitrary, so that any empirical tests
of this expression would require comparisons between the time of measurement
with different couplings $\lambda^{2}$ in order to eliminate the
arbitrary parameter.

\section{Comparisons}

In this section, we compare the upper limit obtained above with cases
where the duration of measurement is shorter. We consider that the
phase noise occurs while the observable $\hat{\sigma}_{x}$ is being
measured. The following numerical results were obtained according
to the superoperator-splitting method described in Ref. \cite{key-14},
to which it is applied the condition from Eq. (\ref{problema}). In
all the simulations, we have chosen the initial state $\frac{1}{\sqrt{2}}\left(\left|+\right\rangle -e^{i\pi/4}\left|-\right\rangle \right)$,
so that the coherences have initially no real part:

\[
\rho_{12}^{\left(x\right)}\left(0\right)=\frac{1}{2}\left(1-e^{i\pi/4}\right)\frac{1}{2}\left(1+e^{-i\pi/4}\right)=-\frac{1}{2}\sin\left(\frac{\pi}{4}\right)i=-\frac{1}{2\sqrt{2}}i.
\]
 Simulations of this part of the coherence are shown in Fig 1.

\begin{figure}
\includegraphics[width=0.5\textwidth]{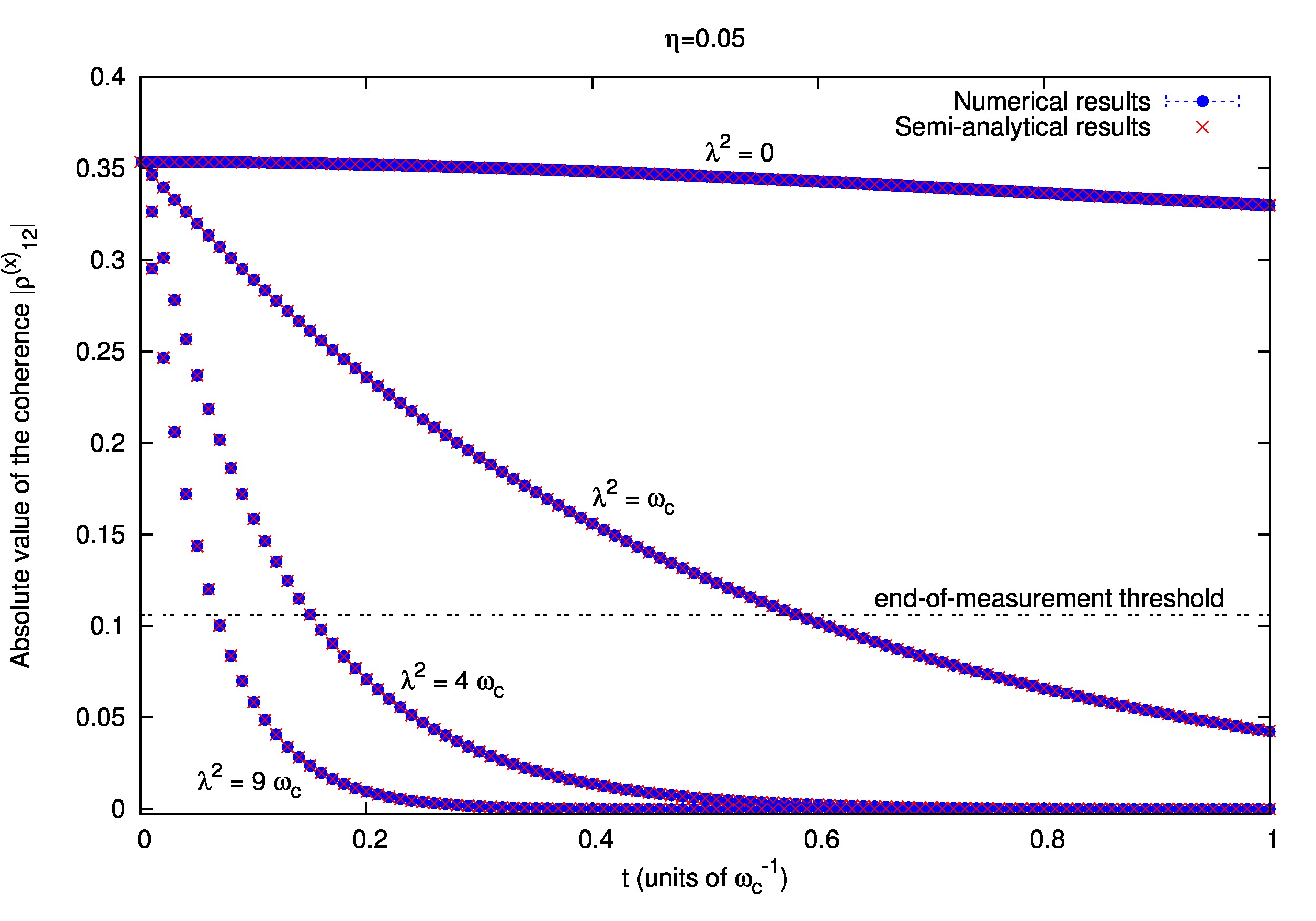}

\includegraphics[width=0.5\textwidth]{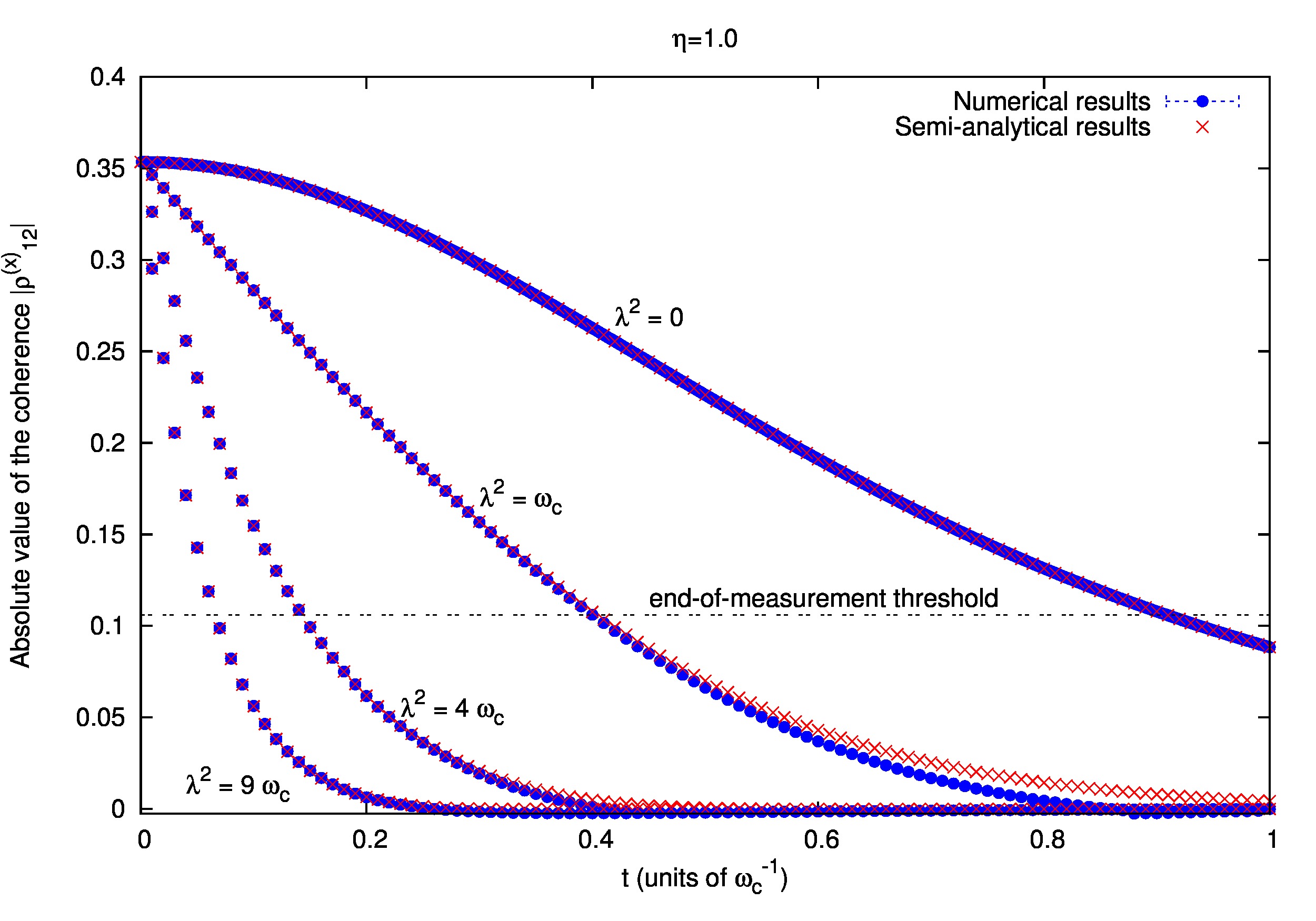}

\includegraphics[width=0.5\textwidth]{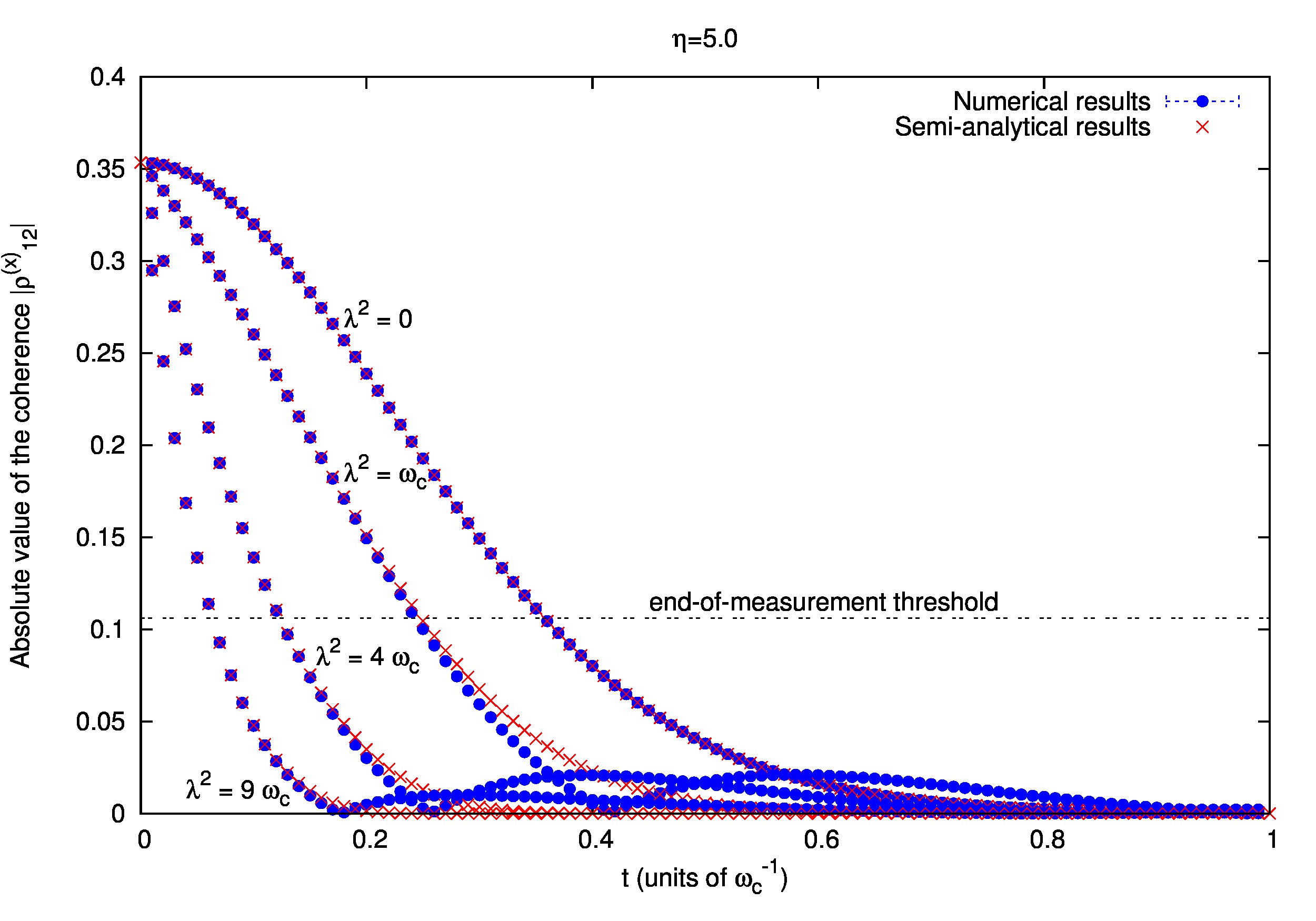}

\caption{(Color online and in black-and-white in print) Time evolution of the
absolute value of the coherence for the initial state $\frac{1}{\sqrt{2}}\left(\left|+\right\rangle -e^{i\pi/4}\left|-\right\rangle \right)$,
for different strengths of measurement ($\lambda$) and both weak
($\eta=0.05$) and strong ($\eta=1,\eta=5$ ) couplings with the environment.
The numerical results are found according to Ref. \cite{key-14},
while the semi-analytical results are those found in Eq. (\ref{roxfinal}),
so that there is a better agreement between the two methods when the
noise is not so intense. Choosing $f=0.3$ to define the end of the
measurement, it can be seen from these curves that the measurement
is faster when the coupling with the apparatus is stronger ($\lambda$
increases) or when the noise is more intense (greater $\eta$).}
\end{figure}

This initial condition requires the simulation of only $\mathrm{Im}\left\{ \rho_{12}^{\left(x\right)}\left(0\right)\right\} $,
which must satisfy

\[
\frac{\mathrm{Im}\left\{ \rho_{12}^{\left(x\right)}\left(t_{M}\right)\right\} }{\mathrm{Im}\left\{ \rho_{12}^{\left(x\right)}\left(0\right)\right\} }\le f
\]
 at the end of the measurement, at instant $t_{M}$. This simplified
condition to assess $t_{M}$ is employed in Fig. 2, where it can be
verified that an increase in $\eta$ or $\lambda$ makes the measurement
process faster.

\begin{figure}
\includegraphics[width=0.5\textwidth]{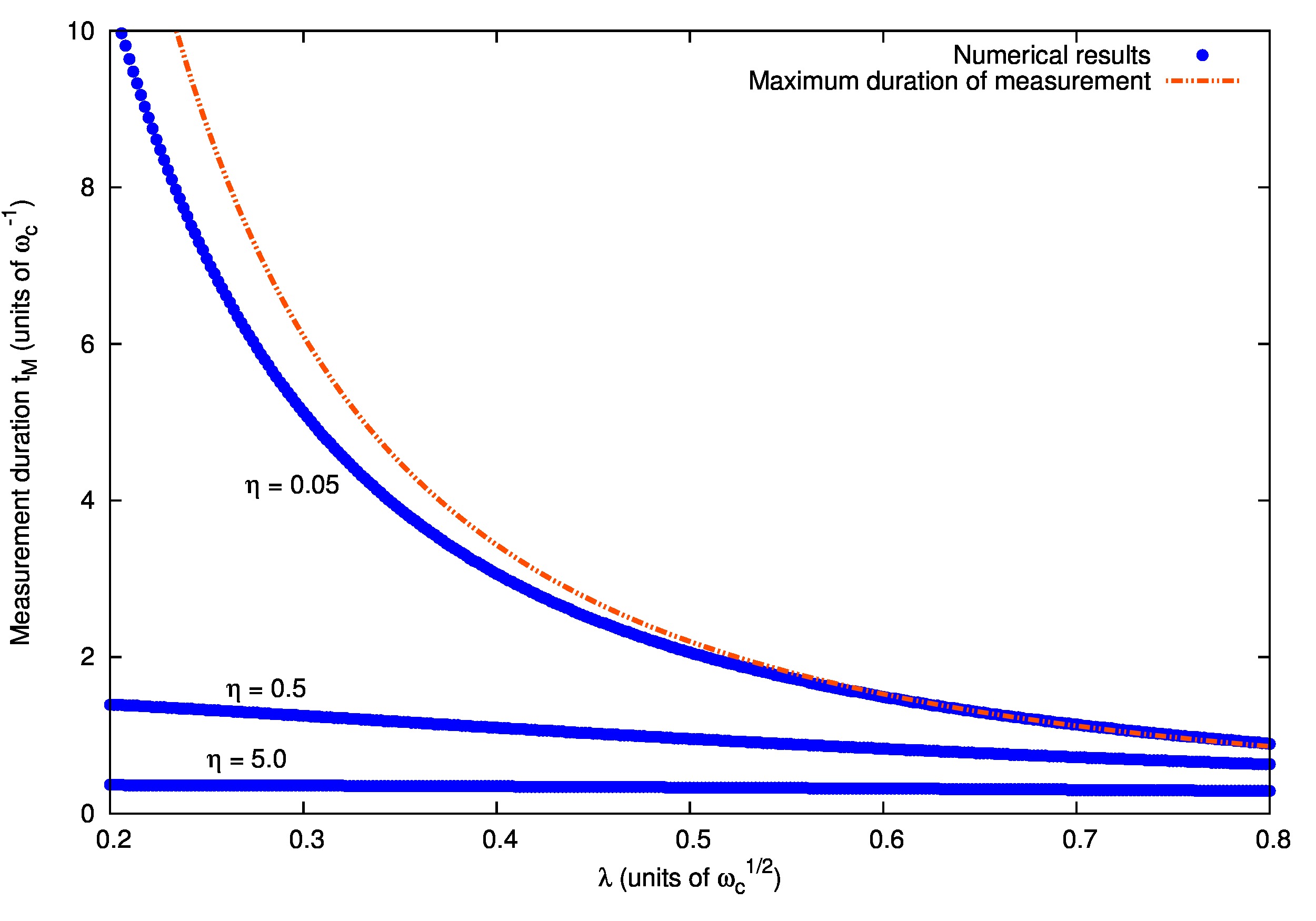}\caption{(Color online and in black-and-white in print) Numerical results for
the duration of the measurement, using $f=0.3$, as a function of
the coupling with the apparatus ($\lambda$), for different values
of $\eta$. The dashed line represents the upper limit to the duration
of measurement, given by Eq. (\ref{tm}). From these curves, it can
be seen once again that both an increase in the noise or in the strength
of measurement can decrease the duration of the measurement process.}
\end{figure}

\section{Conclusions}

Both expressions for the end of the measurement obtained analytically,
Eqs. (\ref{tempoMz}) and (\ref{tempoMx}), are identical. This is
an interesting fact, given that the situations were distinct, not
only because of the different types of measurements being made, but
also because, in the second case, particularizations were made. Qualitatively,
the influence of the environment temperature does not change the conclusions
of this and former works, but the addition of the principal system
behind the $\omega_{0}$ parameters induces some oscillations on the
populations, but the coherences' modulus tends to an asymptotic constant
value, being possible still to use its behavior to define a measurement
time. These new studies will be published soon.

It is interesting to note that Eq. (\ref{tm}) depends only on the
system-measurer coupling $\lambda$ (and, of course, on the fraction
$f$ of the modulus of the coherence we see sufficient to consider
the measurement as completed) and does not depend on initial conditions. 

In a general situation, where $\eta$ is high enough, this parameter
can change the decoherence time. However, here we considered a situation
where the system-environment coupling is small to allow us to expand
the solutions for the coherences in a power series. Then, as expected
intuitively, the influence of $\eta$ is small - more precisely, zero
- and the expression does not depend on the system-environment coupling.
Otherwise, the stronger the system-measurer coupling $\lambda$, the
more intense the measurement, and, consequently, the faster its completion.
On both cases, these intuitive conclusions were rigorously proved.

There are studies on the problem of the \emph{reading }(or \emph{output})
\emph{time} \cite{key-17,key-18}, where this time was considered
as a constant of nature, independent of the system under scrutiny.
In this way, \emph{if} the process of reduction of the wave function
does exists (contrary to the Everettian thesis \cite{key-3,key-4,key-5}),
a complete treatment for the \emph{intervention} problem should include
our measurement time - Eq. (\ref{tm}) - plus the reading time. Anyway,
the approach of this paper does not contradicts the statistical character
of quantum mechanics.

Of course, it all depends on the validity of the measurement time
as proposed in this Lindbladian treatment of the measurement apparatus.
As our expression for the upper limit in the measurement time is not
only simple, but also depends solely on the system-measurer coupling,
it can be empirically tested by varying the strength of the coupling
with the apparatus. Furthermore, our method can be employed in more
complex and, perhaps, more realistic systems (more than two levels,
other types of system-environment interaction, other types of environments,
etc.), with results that can be used in comparative studies against
other quantum-measurement approaches, such as the thermodynamic one
of references\cite{key-20,key-21,key-22}, even though the exact nature
of how this could be accomplished is left for future works.
\begin{acknowledgments}
C. A. Brasil acknowledges support from Coordenação de Aperfeiçoamento
de Pessoal de Nível Superior (CAPES) and Fundação de Amparo à Pesquisa
do Estado de São Paulo (FAPESP), project number 2011/19848-4, Brazil.

L. A. de Castro acknowledges support from FAPESP, project number 2009/12460-0,
Brazil.

R. d. J. Napolitano acknowledges support from Conselho Nacional de
Desenvolvimento Científico e Tecnológico (CNPq), Brazil.\end{acknowledgments}

\end{document}